\documentclass[conference]{IEEEtran}
\IEEEoverridecommandlockouts
\usepackage{cite}
\usepackage{amsmath,amssymb,amsfonts}
\usepackage{algorithmic}
\usepackage{graphicx}
\usepackage{textcomp}
\usepackage{xcolor}
\def\BibTeX{{\rm B\kern-.05em{\sc i\kern-.025em b}\kern-.08em
    T\kern-.1667em\lower.7ex\hbox{E}\kern-.125emX}}
\begin{document}

\title{Demo Paper: A Game Agents Battle Driven by Free-Form Text Commands Using Code-Generation LLM\\}

\author{
    \IEEEauthorblockN{Ray Ito}
    \IEEEauthorblockA{\textit{Faculty of Engineering,} \\
    \textit{The University of Tokyo}\\
    Tokyo, Japan \\
    ray51ito@g.ecc.u-tokyo.ac.jp}
\and
    \IEEEauthorblockN{Junichiro Takahashi}
    \IEEEauthorblockA{\textit{Faculty of Engineering,} \\
    \textit{The University of Tokyo}\\
    Tokyo, Japan \\
    takahashi-junichiro509@g.ecc.u-tokyo.ac.jp}
}

\maketitle

\begin{abstract}
This paper presents a demonstration of our monster battle game, in which the game agents fight in accordance with their player’s language commands. The commands were translated into the knowledge expression called behavior branches by a code-generation large language model. This work facilitated the design of the commanding system more easily, enabling the game agent to comprehend more various and continuous commands than rule-based methods. The results of the commanding and translation process were stored in a database on an Amazon Web Services server for more comprehensive validation. This implementation would provide a sufficient evaluation of this ongoing work, and give insights to the industry that they could use this to develop their interactive game agents.
\end{abstract}

\begin{IEEEkeywords}
Game AI, Character AI, LLM, Game Agent, Knowledge Expression, Human-Computer Interaction, Entertainment Computing
\end{IEEEkeywords}

\section{Introduction}
Game players' desire to control their trained monsters has led to the success of games such as Pokémon, Digimon, and Monster Rancher. Still, players were only able to choose from a limited set of options to command their monsters in these games. This led to attempts, including \cite{b1,b2,b3,b4}, to make game monsters understand and react to language commands, but these implementations were rule-based, which was not easy for developers to hard-code a sufficient interaction. Therefore, \cite{pre} proposed a method to translate player language commands without limitations into the game agent's action. The language command was passed to the code-generation large language model (LLM) model, which generated the knowledge expression called behavior branches. This method enabled the game agent to comprehend a wider range of content and expressions of the player's commands. The concept of using code-generation LLM was originated from \cite{b5}. While \cite{b5} intended to handle independent tasks for robotics, \cite{pre} extended this to the game agents, allowing for the execution of more continuous and rapidly changing commands.

To examine if this method worked, \cite{pre} only implemented a single controllable game agent against a single unmoving agent. However, the effectiveness of this approach in the context of actual player battles remained unexamined. In this paper, we demonstrate a system that enables two players to fight each other at a distance under this commanding system, and the command and translation results are stored in the database on our backend server. Therefore, we aimed to contribute to the advancement of knowledge and entertainment technology in the following ways:

\begin{itemize}
    \item To give practical validation to this ongoing work.
    \item To demonstrate to the industry and researchers that they could use or extend our method to develop their interactive game systems.
\end{itemize}

\section{System Overview}
\subsection{The Entire System}

\begin{figure}[!t]
    \centerline{\includegraphics[width=1.7in]{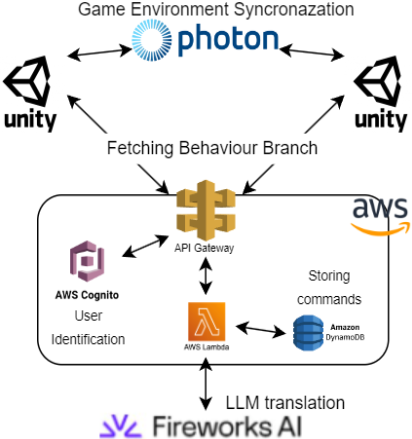}}
    \caption{Overview of the entire system.}
    \label{fig_SystemComponent}
\end{figure}

The system consists of the following components as shown in Fig. \ref{fig_SystemComponent}. Each description is as follows:

\begin{itemize}
    \item \textbf{Unity} (2022.3.15f1): Unity computed the game environment and provided the game environment to the players.
    \item \textbf{Photon PUN2} (2.45): PUN2 was used for the real-time network synchronization between both Unity clients.
    \item \textbf{AWS Server}: The server was used to generate the behavior branches from the player's commands and to store the translation logs. Each players were identified using Cognito, and the logs were stored in DynamoDB. Lambda did the API control including accessing the LLM model. The latency was all together 1.8 seconds on average.
    \item \textbf{Fireworks AI}: The API of Fireworks AI was used for utilizing the default `llama-v2-34b-code' model. Fireworks AI was chosen for its rapid latency. The latency was 0.9 seconds on average.
\end{itemize}

\subsection{Game Environment}

In the game environment, two game agents were placed in the plain 3D space. The game agents could move in the 3D space, and use the following attacks\footnote{These were inspired by Creatures Inc. ``Poképark Wii: Pikachu's Adventure.''}:

\begin{itemize}
    \item \textbf{Thunderbolt}: The game agent shoots a thunderbolt (implemented as a sphere) to the opponent.
    \item \textbf{Iron Tail}: The game agent swings its tail to the opponent.
    \item \textbf{Tackle}: The game agent rushes and hit the opponent.
\end{itemize}

\begin{figure}[!t]
    \centerline{\includegraphics[width=2.5in]{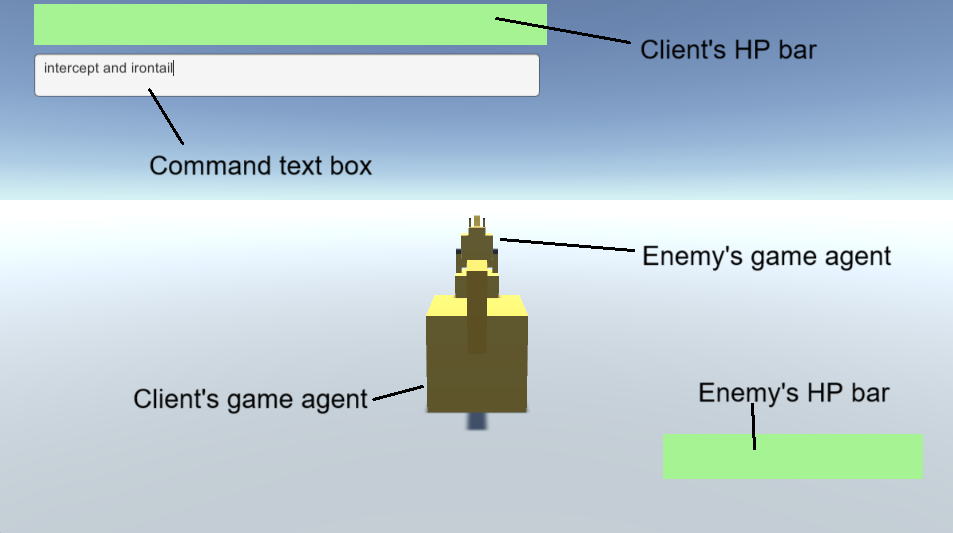}}
    \caption{The visual interface of the game.}
    \label{fig_Interface}
\end{figure}

The interface of the game is shown in Fig. \ref{fig_Interface}. The players could command their game agent by typing the command in the text box. The translation when the enter key was pressed. The game would be paused while the players were typing.

\subsection{Command-Action Translation}
Language commands were translated into the actionable knowledge expression called behavior branches proposed by \cite{pre}. Behavior branches were tree structures consisting of the following nodes:

\begin{itemize}
    \item \textbf{Action node}: Specified the action to be executed.
    \item \textbf{Condition node}: This connected to two nodes, and the satisfaction of this node determined which node to be executed next.
    \item \textbf{Control node}: Specified the controling of the execution flow.
\end{itemize}

This followed the concept of structural programming, and appending different behavior branches could dynamically edit the action flow. As the game agent followed these nodes, \cite{pre} has shown that the game agent could understand and follow the commands freely written under the static experiment\footnote{Static experiment refers that only one game agent was playable and the other stayed still. Thus, the performance under the actual battle, which was the aim of this paper, remained unknown}. For more details of the behavior branches, refer to \cite{pre}.

As similar to \cite{b5}, the prompt for the LLM presented the intended format and the examples of the translations in Python. The inference code was streamed back from the LLM API\footnote{It started to process it as soon as the closing bracket in the generated Python code referring to the behavior branch was detected. This early stopping made the translation 1.1 seconds faster than \cite{pre}.}.

\subsection{Logs in the Database}
\begin{figure}[!t]
    \centerline{\includegraphics[width=3.0in]{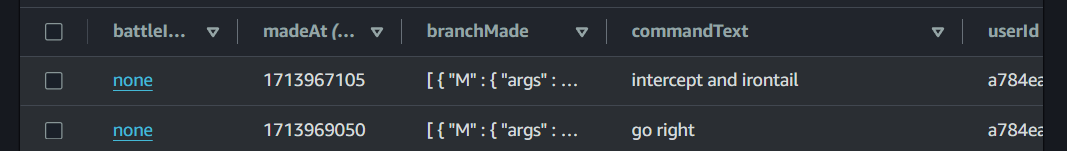}}
    \caption{The screenshot of the table in DynamoDB. Note that any personal information is not shown.}
    \label{fig_DatabaseLogs}
\end{figure}

Fig. \ref{fig_DatabaseLogs} shows a clip of recorded data. There was one row for each command, including the ID of the battle session, the timestamp of the command in Unix time, the player's Cognito ID, the original command text, and the translated behavior branch in JSON format.

\section{Demonstration}
During the demonstration, two attendees will play the game with two laptops prepared. The demonstrator will explain the game rules and what moves the game agent able to do. The attendees will play the game by typing the command in the text box, and try to defeat the opponent's game agent. The game will take approximately 3 minutes. In short, the required facilities for this demo include a table to put the laptops and keyboards, and the internet connection.

\section{Conclusions and Future Works}
In this paper, we demonstrated a battle game system where the game characters flexibly follow the player's directions. The method proposed by \cite{pre}, translating player's commands into Behavior Branches using code-generation LLM, has become available for actual battle gaming. For future work, we plan to conduct a quantitative and qualitative experiment to evaluate using the record and improve our suggested method so as to make it more practical and useful for the game industry.

\end{document}